\documentclass{jps-cp}
\usepackage{txfonts} 
\usepackage{color}

\def\gsim{\mathop {\vtop {\ialign {##\crcr 
$\hfil \displaystyle {>}\hfil $\crcr \noalign {\kern1pt \nointerlineskip } 
$\,\sim$ \crcr \noalign {\kern1pt}}}}\limits}
\def\lsim{\mathop {\vtop {\ialign {##\crcr 
$\hfil \displaystyle {<}\hfil $\crcr \noalign {\kern1pt \nointerlineskip } 
$\,\,\sim$ \crcr \noalign {\kern1pt}}}}\limits}

\title{
Theory for Non-Fermi Liquid Temperature Dependence in Resistivity 
of {Ce$_{x}$La$_{1-x}$Cu$_{5.62}$Au$_{0.38}$ ($x=0.02-0.10$) } on the  
Local Quantum Valence Criticality of Ce Impurities}

\author{Kazumasa \textsc{Miyake}$^{1}$ and Shinji \textsc{Watanabe}$^{2}$}

\inst{
$^{1}$
Center for Advanced High Magnetic Field Science, Osaka University, Toyonaka, 
560-0043, Japan \\
$^{2}$
Department of Basic Sciences, Kyushu Institute of Technology, Kitakyushu, 
804-8550, Japan
}

\email{
miyake@mp.es.osaka-u.ac.jp
}

\recdate{September 9, 2019}

\abst{
It was reported by Shiino {\it et al} in J. Phys. Soc. Jpn. {\bf 86}, 123705 (2017) that 
Ce$_{x}$La$_{1-x}$Cu$_{5.62}$Au$_{0.38}$ ($x=0.02-0.10$) 
exhibits a new type of quantum criticality in both magnetic and 
thermal properties, which is the same as that observed in a series of materials exhibiting 
quantum critical valence fluctuations (QCVF), such as $\beta$-YbAlB$_4$, Yb$_{15}$Al$_{34}$Au$_{51}$, 
and so on.  However, the temperature ($T$) dependence in the resistivity $\rho(T)$ for $T<0.5\,$K 
is quite anomalous, i.e., $\rho(T)\propto ({\rm const.}-T^{n})$ with $n\simeq 0.75$ {at $x=0.05$}. 
We find that this anomalous exponent  
is given by $n=2(1-\zeta)$, with $\zeta$ being the weakly temperature dependent 
($0.5\lsim \zeta \lsim 0.7$) critical exponent for the QCVF. 
The observed critical exponent $n\simeq 0.75$ at $x=0.05$ is reproduced by 
choosing $\zeta\simeq 0.63$ 
which is consistent with the divergent behavior observed in the uniform magnetic susceptibility 
$\chi\propto T^{-\zeta}$ with $\zeta\simeq 0.67$ at $x=0.02$.
}

\kword{local quantum valence criticality, 
Ce$_{x}$La$_{1-x}$Cu$_{5.62}$Au$_{0.38}$ ($x=0.02-0.10$), 
non-Fermi liquid resistivity
}

\begin{document}
\maketitle

\section{Introduction}
It has been pointed out 
in Refs.\ [\citen{Miyake1,Miyake2}] that CeCu$_6$ exhibits a rather sharp crossover 
in the valence of Ce ion under pressure~\cite{Raymond} though it is a bit milder than the case of 
CeCu$_2$(Si,Ge)$_2$~\cite{Jaccard,Holmes}.  
Recently, a symptom of magnetic field and pressure 
induced quantum critical valence transition (QCVT) was reported~\cite{Hirose,MiyakePM}, which had been 
predicted in Refs.\ [\citen{Watanabe3,Watanabe3a}]. 
In this sense, it may be reasonable to explore a possibility that the QCVT is realized if the 
magnetic order in CeCu$_{6-x}$Au$_{x}$ ($x\gsim0.1$)   
is suppressed by partly replacing Ce by 
La because the QCVT is an essentially local phenomenon~\cite{Onishi1,Watanabe1}.  
Indeed, recently, it was reported in Ref.\ [\citen{Shiino}] that 
Ce$_{x}$La$_{1-x}$Cu$_{5.62}$Au$_{0.38}$ ($x=0.02-0.10$) 
exhibits non-Fermi liquid properties in temperature ($T$) dependence of the magnetic susceptibility and 
the specific heat, and the $T/B$ scaling of the magnetization, which are the same as those expected 
at around the QCVT in periodic~\cite{Watanabe1} or aperiodic lattice 
systems~\cite{Watanabe2,Watanabe5}. 

While one might think that the system with $x=0.02$ can be described by the single impurity model, 
the mean distance between Ce ions is only about 4$\times$(lattice constant) which is not large enough 
compared to the size of cloud of the Kondo-Yosida singlet state discussed in Ref.\ [\citen{Yosida-Yoshimori}],  
so that the collective effect of Ce impurities should be taken into account. 
In other words, the theory for QCVT can be applied through averaging process over 
the random distribution of Ce ions. 
It is also reasonable to expect as noted above \cite{Miyake1,Miyake2} that the quantum criticality of 
valence transition is induced by diluting Ce by La which has a larger atomic or ionic radius than Ce 
as reported in, e.g., Ref. [\citen{Satoh}], because Ce ions feel positive pressure locally 
from surrounding and expanding medium which includes La ions with larger ionic radius, 
if the concentration of Ce is dilute enough. 
This positive pressure effect changes the system from the Kondo regime  to the valence fluctuations one 
with higher Kondo temperature, where the sharp valence crossover occurs as reported in 
Ref.\ [\citen{Raymond}]. 

On the other hand, the resistivity $\rho(T)$ exhibits the $T$ dependence like 
$\rho(T)\propto ({\rm const.}-T^{n})$ with $n\simeq 0.75$ which is quite anomalous even from 
the view point of the QCVT in the lattice systems.  The purpose of the present paper is to derive this 
anomalous $T$ dependence on the basis of the QCVT scenario.

\section{Model Hamiltonian}
A canonical model for describing the valence transition is the 
extended periodic Anderson model that take{s} into account 
the Coulomb repulsion $U_{\rm fc}$ between f and conduction 
electrons~\cite{Onishi1,Watanabe1,Watanabe2,Watanabe5}. 
To discuss the present situation, 
we have to revise the model as follows:    
\begin{eqnarray}
& &
H_{\rm EPAM}
 = \sum_{{\bf k},\sigma}\epsilon_{\bf k}c_{{\bf k}\sigma}^{\dagger}c_{{\bf k}\sigma}
 +\varepsilon_{\rm f} \sum_{i,\sigma}f_{i\sigma}^{\dagger}f_{i\sigma}
+\frac{1}{\sqrt{N_{\rm L}}}\sum_{i,{\bf k},\sigma}\left(V_{\bf k}e^{{\rm i}{\bf k}\cdot{\bf r}_i}
c_{{\bf k}\sigma}^{\dagger}f_{i\sigma}^{}+{\rm h.c.}\right)
\nonumber
\\
& &
\qquad\qquad\qquad
+U_{\rm ff}\sum_i n_{i \uparrow}^{\rm f} n_{i \downarrow}^{\rm f}
+U_{\rm fc}\sum_{i,\sigma \sigma'}n_{i \sigma}^{\rm f}n_{i \sigma'}^{\rm c},
 \label{EPAM}
\end{eqnarray}
where {f electrons occupy the $i$-sites which are randomly distributed in diluted system, 
and conduction electrons are described by wave vector ${\bf k}$ 
defined on the periodic lattice sites of $N_{\rm L}$  
and have energy dispersion $\epsilon_{\bf k}$.  }
The label $\sigma$ in Eq.\ (\ref{EPAM}) stands for the Kramers doublet state of the ground Crystalline-Electric-Field (CEF) level.   
Hereafter, the c-f hybridization $V_{{\bf k},i}$ at $i$-site is assumed to be constant $V$. 

The {type of} 
Hamiltonian [Eq.\ (\ref{EPAM})] well describes phenomena associated with quantum critical valence 
fluctuations (QCVF) not only in periodic lattice systems such as 
$\beta$-YbAlB$_4$~\cite{Watanabe1,Nakatsuji,Matsumoto}  
but also in aperiodic systems, such as clusters forming Tsai-type quasicrystal 
Yb$_{15}$Al$_{34}$Au$_{51}$~\cite{Watanabe2,Deguchi}, and approximant 
Yb$_{14}$Al$_{35}$Au$_{51}$~\cite{Watanabe5,Matsukawa}, all of which exhibit the unconventional 
non-Fermi liquid behaviors of the same universality class 
different from those in the quantum criticality associated with itinerant magnetic 
transitions~\cite{Moriya}.    
The point is that the periodicity is not essential for such unconventional universality class to appear 
because the QCVT is quite local in character.

\section{Recipe for Average over Random Distribution of Ce Ions}

A basic idea for taking the effect of Ce ions in diluted system is that the 4f electrons at Ce sites 
acquire the wave 
vector ${\bf q}$ dependence through the average over the impurities distribution.   
By this process, there arises two contributions to the scattering process of the 
conduction electrons, i.e., a single site effect of localized f electrons and the lattice effect 
due to the wave-number dependent collective valence fluctuations (VF).   
Hereafter, 
we consider the case in which the concentration $c_{\rm imp}\equiv N_{\rm f}/N_{\rm L}$ ($x$) of Ce ions 
in Ce$_{x}$La$_{1-x}$Cu$_{5.62}$Au$_{0.38}$ is considerably smaller than 1 but is non-vanishing 
in the thermodynamic limit, $N_{\rm f}\to \infty$ and $N_{\rm L}\to \infty$. 

\begin{figure}[h]
\begin{center}
\rotatebox{0}{\includegraphics[width=0.7\linewidth]{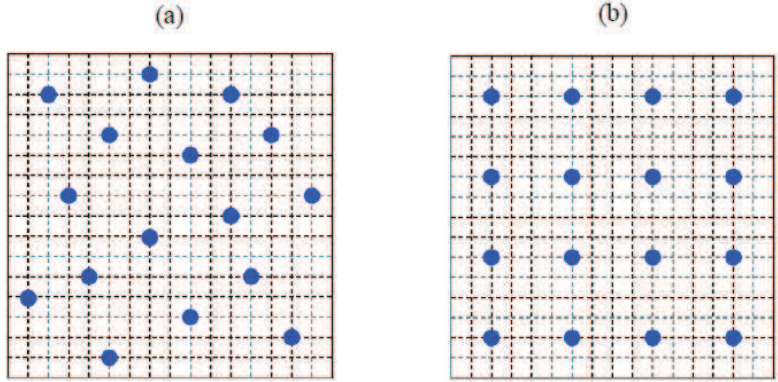}}
\caption{
(a) Square lattice version of original system where the sites occupied by Ce ions are shown 
by filled circle. (b) Virtual system obtained after average over random distribution of Ce sites. 
}
\label{Fig:Replica}
\end{center}
\end{figure}

To estimate the effect of scattering due to the random distribution of Ce ions, 
we have to take an average over the distribution.  
Before taking the average, the one-particle Green function of f electron depends on two positions 
as $G_{\rm f}({\bf r}_{i}, \tau;{\bf r}_{j}, \tau^{\prime})$, which 
becomes a function of the relative coordinate $({\bf r}_{i}-{\bf r}_{j})$ after the average 
as in the usual case discussing impurity scattering effect~{\cite{AGD}}. 
Namely, by taking this average, the wave-vector dependent Green function is defined as 
\begin{eqnarray}
{\tilde G}_{{\rm {\tilde f}}}({\bf p},\tau-\tau^{\prime})
\equiv\frac{1}{N_{\rm f}}\sum_{(i-j)}e^{-{\rm i}{\bf p}\cdot({\bf r}_{i}-{\bf r}_{j})}
\langle\langle G_{\rm f}({\bf r}_{i},{\bf r}_{j}; \tau-\tau^{\prime})\rangle\rangle, 
\label{eq:Green_f1}
\end{eqnarray}
where $\langle\langle\cdots\rangle\rangle$ denotes the average over the random distribution of Ce ions, 
and $N_{\rm f}$ is the number of {\it virtual} lattice sites 
occupied by f electrons as schematically shown in Fig.\ \ref{Fig:Replica}(b).   
Note here that the $N_{\rm f}$ appears 
instead of the number $N_{\rm L}$ of the original lattice sites. In other words, the lattice constant 
of the {\it virtual} {periodic} lattice system of Ce is enlarged by a factor 
{$c_{\rm imp}^{-1/3}$.} 

Hereafter, ${\bf p}$'s are used for the wave vector in the {\it virtual} lattice obtained after 
the random average of Ce ions, and distinguished from wave vectors ${\bf k}$'s in the original lattice 
shown in Fig.\ \ref{Fig:Replica}(a).  

On the other hand, conduction electrons described by wave vector ${\bf p}$ are essentially 
unaltered by the effect of scattering by Ce impurities  except for some broadening of the 
dispersion due to impurities scattering~\cite{AGD}.  Namely, for example, the density of states (DOS) of 
conduction electrons at the Fermi level are essentially unaltered.   
However, since the size of the Brillouin zone (BZ) of the {\it virtual} lattice is shortened 
by a factor $c_{\rm imp}^{1/3}$, 
the band of conduction electrons splits into multibands in the shortened and reduced BZ.  
However, we use an extended zone scheme for conduction electrons.

Thus, the {\it virtual} Hamiltonian $H^{\rm v}_{\rm EPAM}$ is given explicitly as follows:
\begin{eqnarray}
& &
H^{\rm v}_{\rm EPAM}\equiv \sum_{{\bf p},\sigma}
{\epsilon}_{{\bf p}}{\tilde c}_{{\bf p}\sigma}^{\dagger}
{\tilde c}_{{\bf p}\sigma}
+\sum_{{i},\sigma}u_{\rm imp}P_{i}\,n_{i}^{{\tilde c}}
 +\varepsilon_{\rm f} \sum_{{{\tilde i}},\sigma}{{\tilde f}}_{{{\tilde i}}
\sigma}^{\dagger}
 {{\tilde f}}_{{{\tilde i}}\sigma}
 +\sqrt{c_{\rm imp}}\sum_{{\bf p},\sigma}(
V{\tilde c}_{{\bf p}\sigma}^{\dagger}
{{\tilde f}_{{\tilde i}\sigma}}+{\rm h.c.})
 \nonumber
 \\
 & &
 \qquad\qquad\qquad
+\,U_{\rm ff}\sum_{{\tilde i}} n_{{{\tilde i}} \uparrow}^{\rm {{\tilde f}}} 
n_{{{\tilde i}} \downarrow}^{\rm {{\tilde f}}} 
+U_{\rm fc}\sum_{{{\tilde i}},\sigma \sigma'}
n_{{{\tilde i}} \sigma}^{\rm {{\tilde f}}}
n_{{{\tilde i}} \sigma'}^{\rm {{\tilde c}}},
 \label{eq:Replica}
\end{eqnarray}
where 
$\tilde{c}_{{\bf p}\sigma}$ and $\tilde{f}_{{\bf p}\sigma}$ are the annihilation operators 
of conduction- and f-electrons in the {\it virtual} periodic lattice system.  
Note that the factor $\sqrt{c_{\rm imp}}\equiv\sqrt{N_{\rm f}/N_{\rm L}}$ in the fourth term of 
Eq.\ (\ref{eq:Replica}) reflects the fact that 
the f electrons are located on the periodic {\it virtual} lattice points of $N_{\rm f}$ 
while the conduction electrons are hopping among the original lattice points of $N_{\rm L}$ 
including the points $i\not={\tilde i}$, as discussed in detail elsewhere. 
This factor manifests the diluteness of Ce ions in the original lattice [Fig.\ \ref{Fig:Replica}(a)].  
The random variables {$P_{i}$} in the second term of Eq.\ (\ref{eq:Replica}) represent 
the effect of impurity scattering in the original lattice shown in Fig.\ \ref{Fig:Replica}(a). 
With the use of this replica Hamiltonian,  
the theoretical framework discussing QCVF can be applied as it is because the QCVF are essentially 
the local phenomena~\cite{Watanabe1}. 

Since the the {\it virtual} Hamiltonian [Eq.\ (\ref{eq:Replica})] is essentially the same as that for the 
periodic lattice system (except the random variable and the factor $\sqrt{c_{\rm imp}}$ weakening 
the c-f hybridization), the condition for the  QCVT is not altered seriously as will be discussed 
more quantitatively elsewhere.   
This is because the valence transition is essentially the {\it local} phenomenon or valence susceptibility 
has extremely weak wave-vector dependence \cite{Onishi1, Watanabe1}.

\section{Two Contributions to Resistivity}
After taking  the average over 
the random distribution of Ce ions, both the Green function and the self-energy of conduction electrons 
become the function of the 
relative coordinate $({\bf r}_{i}-{\bf r}_{j})$'s and acquire the wave vector representation.  
Namely, the Green function of conductin electrons defined as 
${\tilde G}_{\rm {\tilde c}}({\bf p},{\rm i}\varepsilon_{n})\equiv
(1/N_{\rm L})\sum_{(i-j)}e^{-{\rm i}{\bf p}\cdot({\bf r}_{i}-{\bf r}_{j})}
\langle\langle G_{\rm c}({\bf r}_{i},{\bf r}_{j},{\rm i}\varepsilon_{n})\rangle\rangle$ satisfies the 
Dyson equation as 
\begin{eqnarray}
& &
{\tilde G}_{\rm {\tilde c}}({\bf p},{\rm i}\varepsilon_{n})=
G^{(0)}_{\rm c}({\bf p},{\rm i}\varepsilon_{n})
+G^{(0)}_{\rm c}({\bf p},{\rm i}\varepsilon_{n}){\tilde \Sigma}_{\rm {\tilde c}}({\bf p},{\rm i}\varepsilon_{n})
{\tilde G}_{\rm {\tilde c}}({\bf p},{\rm i}\varepsilon_{n}),
\label{Green_c_p}
\end{eqnarray}
where $G^{(0)}_{\rm c}({\bf p},{\rm i}\varepsilon_{n})$ is the Green function of free band electrons 
on the original lattice, and ${\tilde \Sigma}_{\rm {\tilde c}}({\bf p},{\rm i}\varepsilon_{n})$ is defined by 
\begin{eqnarray}
& &
{\tilde \Sigma}_{\rm {\tilde c}}({\bf p},{\rm i}\varepsilon_{n})\equiv
\frac{1}{N_{\rm L}}\sum_{(i-j)}e^{-{\rm i}{\bf p}\cdot({\bf r}_{i}-{\bf r}_{j})}
\langle\langle \Sigma_{\rm c}({\bf r}_{i},{\bf r}_{j},{\rm i}\varepsilon_{n})\rangle\rangle.
\label{Sigma_c_p}
\end{eqnarray}
Note that the conduction electrons are defined on the original lattice points as shown in 
Fig.\ \ref{Fig:Replica}(a) so that the factor $(1/N_{\rm L})$ appears in Eq.\ (\ref{Sigma_c_p}) 
instead of  the factor $(1/N_{\rm f})$ appearing in the definition of 
${\tilde G}_{\rm {\tilde f}}({\bf p},\tau-\tau^{\prime})$ [Eq.\ (\ref{eq:Green_f1})].  

The $T$ dependence of the resistivity is essentially given by the imaginary part of the 
retarded function of ${\tilde \Sigma}_{\tilde c}({\bf p},{\rm i}\varepsilon_{n})$ [Eq.\ (\ref{Sigma_c_p})], 
in which $\langle\langle \Sigma_{\rm c}({\bf r}_{i},{\bf r}_{j},{\rm i}\varepsilon_{n})\rangle\rangle$ 
consists of two parts as 
\begin{eqnarray}
& &
\langle\langle \Sigma_{\rm c}({\bf r}_{i},{\bf r}_{j},{\rm i}\varepsilon_{n})\rangle\rangle
=\delta_{i,\ell}\delta_{j,\ell}
\langle\langle \Sigma_{\rm c}({\bf r}_{\ell},{\bf r}_{\ell},{\rm i}\varepsilon_{n})\rangle\rangle
+\langle\langle\Delta\Sigma_{\rm c}({\bf r}_{i},{\bf r}_{j},{\rm i}\varepsilon_{n})\rangle\rangle,
\label{Sigma_c_p:2}
\end{eqnarray}
where ${\bf r}_{\ell}$ is the site of Ce ions and 
$\langle\langle \Sigma_{\rm c}({\bf r}_{\ell},{\bf r}_{\ell},{\rm i}\varepsilon_{n})\rangle\rangle$ is 
independent of ${\bf r}_{\ell}$, and the term with $i=j=\ell$ is excluded in 
$\langle\langle\Delta\Sigma_{\rm c}({\bf r}_{i},{\bf r}_{j},{\rm i}\varepsilon_{n})\rangle\rangle$. 
The first part in Eq.\ (\ref{Sigma_c_p:2}) represents the damping effect arising from 
independent but {\it dynamical} scattering by localized f {electron at site ${\bf r}_{\ell}$}, 
and the second part arises from the scattering 
by the collective VF described by the {\it virtual} Hamiltonian [Eq.\ (\ref{eq:Replica})].  
Namely, the Green function of conduction electrons 
${\tilde G}_{\rm {\tilde c}}({\bf p},{\rm i}\varepsilon_{n})$ has the following structure: 
\begin{eqnarray}
& &
{\tilde G}_{\rm {\tilde c}}({\bf p},{\rm i}\varepsilon_{n})=
\left[
{\rm i}\varepsilon_{n}-\xi_{\bf p}-c_{\rm imp}
\langle\langle \Sigma_{\rm c}({\bf r}_{\ell},{\bf r}_{\ell},{\rm i}\varepsilon_{n})\rangle\rangle
-\langle\langle\Delta\Sigma_{\rm c}({\bf p},{\rm i}\varepsilon_{n})\rangle\rangle
\right]^{-1},
\label{Green_c_p:2}
\end{eqnarray}
where $c_{\rm imp}$ is the concentration of Ce ions on the original lattice shown in Fig.\ \ref{Fig:Replica}(a), 
and $\langle\langle\Delta\Sigma_{\rm c}({\bf p},{\rm i}\varepsilon_{n})\rangle\rangle$ is the wave-vector 
representation of 
$\langle\langle\Delta\Sigma_{\rm c}({\bf r}_{i},{\bf r}_{j},{\rm i}\varepsilon_{n})\rangle\rangle$ defined by 
\begin{eqnarray}
\langle\langle\Delta\Sigma_{\rm c}({\bf p},{\rm i}\varepsilon_{n})\rangle\rangle
\equiv \frac{1}{N_{\rm L}}
\sum_{(i-j)}e^{-{\rm i}{\bf p}\cdot({\bf r}_{i}-{\bf r}_{j})}
\langle\langle\Delta\Sigma_{\rm c}({\bf r}_{i},{\bf r}_{j},{\rm i}\varepsilon_{n})\rangle\rangle.
\label{Self-energy_collective}
\end{eqnarray}
Here, it is crucial to note that 
$\langle\langle\Delta\Sigma_{\rm c}({\bf r}_{i},{\bf r}_{j},{\rm i}\varepsilon_{n})\rangle\rangle$ can be 
replaced by that of 
$\langle\langle\Sigma_{\rm c}({\bf r}_{i},{\bf r}_{j},{\rm i}\varepsilon_{n})\rangle\rangle$ described by the 
{\it virtual} Hamiltonian [Eq.\ (\ref{eq:Replica})] in which the terms with $i=j=\ell$ are retained. 
This is because the difference arising from the inclusion of this term gives only the effect of the order 
of ${\cal O}(1/N_{\rm f})$ which is negligible in the bulk limit. 
This contribution gives the same $T$ dependence as the resistivity of lattice system, 
$\rho_{\rm lattice}(T) \propto T$, except the factor $c_{\rm imp}$ 
arising from the scaling of the hybridization, from $V$ to $\sqrt{c_{\rm imp}}\,V$, 
which is consistent with the physical picture that the conduction electrons are scattered by 
collective VF of f electrons with concentration $c_{\rm imp}$.   
This $T$ dependence is weaker than that arising from 
$\langle\langle \Sigma_{\rm c}({\bf r}_{\ell},{\bf r}_{\ell},{\rm i}\varepsilon_{n})\rangle\rangle$ 
in Eq.\ (\ref{Green_c_p:2}) 
as discussed in the next section, so that it can be neglected in the low $T$ limit.

\section{Resistivity $\rho_{\rm imp}$ from Incoherent Scattering by Localized f Electrons} 
\begin{figure}[h]
\begin{center}
\rotatebox{0}{\includegraphics[width=0.6\linewidth]{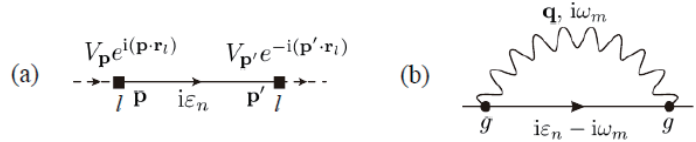}}
\caption{
The Feynman diagrams for (a)  
$\langle\langle \Sigma^{\rm R}_{\rm c}({\bf r}_{\ell},{\bf r}_{\ell},{\rm i}\varepsilon_{n})\rangle\rangle$ 
and (b) the selfenergy due to the VF. 
The solid and dashed lines represent the Green functions of f renormalized only by local correlations 
and conduction electrons at the Ce site ${\bf r}_{\ell}$, respectively.
The squares represent the c-f hybridization at the site ${\bf r}_{\ell}$. 
so that the summation over ${\bf p}$ and ${\bf p}^{\prime}$ should be taken. 
The wavy line represents the VF propagator $\chi_{\rm v}$ [Eq.\ (\ref{chi:1})]  
and $g$ the coupling constant between 
the localized 4f electron and the VF mode, respectively. 
}
\label{Fig:0}
\end{center}
\end{figure}

In this section, we derive the $T$ dependence in the resistivity $\rho_{\rm imp}(T)$ which is given 
with the use of  the retarded function 
$-c_{\rm imp}{\rm Im}
\langle\langle \Sigma^{\rm R}_{\rm c}({\bf r}_{\ell},{\bf r}_{\ell},\varepsilon+{\rm i}\delta)\rangle\rangle$ 
in Eq.\ (\ref{Green_c_p:2}). Then, the leading $T$ dependence of $\rho_{\rm imp}(T)$
is proportional to ${\rm Im}
\langle\langle \Sigma^{\rm R}_{\rm c}({\bf r}_{\ell},{\bf r}_{\ell},{\rm i}\delta)\rangle\rangle$ 
as 
\begin{equation}
\rho_{\rm imp}(T) \propto -c_{\rm imp}{\rm Im}
\langle\langle \Sigma^{\rm R}_{\rm c}({\bf r}_{\ell},{\bf r}_{\ell},{\rm i}\delta)\rangle\rangle,
\label{eq:3}
\end{equation}
because 
${\rm Im}\langle\langle \Sigma^{\rm R}_{\rm c}({\bf r}_{\ell},{\bf r}_{\ell},\varepsilon+{\rm i}\delta)
\rangle\rangle$ 
is proportional to $({\rm const.}-\varepsilon^{2})$ 
in the region, $\varepsilon\sim 0$, 
so that it gives only a conventional $T$ dependence as 
$({\rm const.}-T^{2})$ of the conventional local Fermi liquid. 
The Feynman diagram for 
$\langle\langle \Sigma^{\rm R}_{\rm c}({\bf r}_{\ell},{\bf r}_{\ell},\varepsilon+{\rm i}\delta)\rangle\rangle$ 
is given by Fig.\ \ref{Fig:0}(a), leading to  
\begin{eqnarray}
& &
\langle\langle \Sigma_{\rm c}^{\rm R}({\bf r}_{\ell},{\bf r}_{\ell},\varepsilon+{\rm i}\delta)\rangle\rangle
=\frac{V^{2}}{N_{\rm f}}\sum_{{\bf p}}
{\tilde G}_{\rm {\tilde f}}^{\rm R}({\bf p},\varepsilon+{\rm i}\delta)
\equiv
V^{2}{\langle\langle}{\tilde G}_{\tilde{\rm f}}^{\rm R}(\varepsilon+{\rm i}\delta)\rangle\rangle.
\label{eq:4}
\end{eqnarray}

The Green function 
${\langle\langle}{\tilde G}_{\tilde{\rm f}}^{\rm R}(\varepsilon+{\rm i}\delta)\rangle\rangle$ 
in Eq.\ (\ref{eq:4}) is expressed as
\begin{equation}
{\langle\langle} {\tilde G}_{\tilde{\rm f}}^{\rm R}(\varepsilon+{\rm i}\delta){\rangle\rangle}^{-1}
=[{\bar G}_{\rm f}^{\rm R}(\varepsilon+{\rm i}\delta)]^{-1}
-{\langle\langle}\Sigma_{\tilde{\rm f}}^{\rm R}(\varepsilon+{\rm i}\delta)
{\rangle\rangle}.
\label{eq:5}
\end{equation}
where 
${\bar G}_{\tilde{\rm f}}(\varepsilon)\equiv \bar {z}_{\rm f}/(\varepsilon+{\rm i}{\bar \Delta})$ 
is the Green function of a localized f electron which is 
renormalized {\it only} by local correlation due to $U_{\rm ff}$. 
Here, ${\bar z}_{\rm f}$ is the renormalization amplitude and 
${\bar \Delta}\equiv {\bar z}_{\rm f}\pi V^{2}\rangle N_{\rm cF}$, with 
${N_{\rm cF}}$ being the DOS 
of conduction electrons per electron at the Fermi level,  is the broadening width 
of the local 4f level due to the c-f hybridization effect.  
Then, $\rho_{\rm imp}(T)$ [Eq.\ (\ref{eq:3})] is given as follows: 
\begin{equation}
\rho_{\rm imp}(T)
\propto c_{\rm imp}V^{2}\left[\frac{{\bar \Delta}}{{\bar z}_{\rm f}}-{\rm Im}\,
{\langle\langle}\Sigma^{\rm R}_{{\tilde{\rm f}}}({\rm i}\delta){\rangle\rangle}\right]^{-1}. 
\label{eq:7}
\end{equation}
Since ${\bar \Delta}/{\bar z}_{\rm f}=\pi V^{2}N_{\rm cF}$ 
is essentially $T$ independent in the wide region $T\lsim T_{\rm F}^{*}\ll D$, 
the $T$ dependence of  
$\rho_{\rm imp}(T)$ in the region $T\ll T_{\rm F}^{*}$ arises from that of 
${\rm Im}\,{\langle\langle}\Sigma^{\rm R}_{\tilde{\rm f}}({\rm i}\delta){\rangle\rangle}$. 

The selfenergy 
${\langle\langle}\Sigma^{\rm R}_{\tilde{\rm f}}(\varepsilon+{\rm i}\delta){\rangle\rangle}$ 
of the localized ${\tilde {\rm f}}$ electron (averaged over 
the Ce sites) in Eq.\ (\ref{eq:5}) represents the contribution 
from the effect of the local VF.   
Its explicit form in Matsubara frequency representation 
is given by the Feynman 
diagram shown in Fig.\ \ref{Fig:0}(b), and its analytic form is given by
\begin{equation}
{\langle\langle}\Sigma_{\tilde{\rm f}}({\rm i}\varepsilon_{n})
{\rangle\rangle}=g^{2}T\sum_{\omega_{m}}{\frac{1}{N_{\rm f}}}\sum_{{\bf q}}
\chi_{\rm v}({\bf q},{\rm i}\omega_{m})
{\bar G}_{\rm {{\tilde f}}}({\rm i}\varepsilon_{n}-{\rm i}\omega_{m}), 
\label{eq:8}
\end{equation}
where $\chi_{\rm v}$ is the VF propagator {described by the 
{\it virtual} Hamiltonian [Eq.\ (\ref{eq:Replica})],} and $g$ is the coupling constant between 
the localized 4f electron and the VF mode.  

As discussed in Ref. \cite{Watanabe1} and argued in the Introduction, 
the explicit form of $\chi_{\rm v}({\bf q},{\rm i}\omega_{m})$ in {Eq.\ (\ref{eq:8})} is given 
in parallel to the lattice system as 
\begin{equation}
\chi_{\rm v}({\bf q},{\rm i}\omega_{m})=\frac{N_{\rm F}^{*}}{\eta+Aq^{2}+C|\omega_{m}|}, 
\label{chi:1}
\end{equation}
where $N_{\rm F}^{*}$ is the DOS of quasiparticles renormalized by the 
on-site correlation effect  
arising from $U_{\rm ff}$ in Eq.\ (\ref{EPAM}), the dimensionless parameter $\eta$ ($\lsim 1$ in general)  
is the measure of the distance from the criticality ($\eta=0$), and the coefficients $A$ and $C$ 
parameterize the extent of QCVF in space and time, respectively. 
As discussed in the end of Sect.\ 3, the condition for the QCVT in the present impurities system 
is essentially given by that in the lattice system 
because of the {\it locality} of QCVT, which is manifested itself as an extremely small coefficient $A$ in 
$\chi_{\rm v}({\bf q},{\rm i}\omega_{m})$ [Eq.\ (\ref{chi:1})] leading to extremely small energy 
scale VF, $T_{0}\equiv Aq_{\rm c}^{2}/2\pi C$ ($q_{\rm c}$ being the wave vector cutoff), 
as shown in Ref.~\cite{Watanabe1}. 

With the use of the fact that the relations, $\Gamma_{q}\ll {\bar \Delta}$ and $T\ll {\bar \Delta}$, 
hold near the quantum criticality,  
${\rm Im}\,\Sigma_{\rm f}^{\rm R}({\rm i}\delta)$ derived from Eq.\ (\ref{eq:8}) 
is approximated as  
\begin{equation}
{\rm Im}\,\Sigma^{\rm R}_{\tilde{\rm f}}({\rm i}\delta)
\approx
-\frac{2g^{2}}{\pi}\frac{{\bar z}_{\rm f}N_{\rm F}^{*}}
{C{\bar \Delta}{N_{\rm f}}}
\sum_{{\bf q}}\left[\Phi\left(\frac{\Gamma_{q}}{2\pi T}\right)-\Phi\left(\frac{\Gamma_{q}}{\pi T}\right)\right].
\label{eq:16}
\end{equation}
where $\Phi(y)\equiv \log\,y -({1}/{2y})-\psi(y)$, with $\psi(y)$ being the digamma function, and 
$\Gamma_{\bf q}\equiv(\eta+Aq^{2})/C$. 
In the wide $T$ region,  $T_{0}<T\ll T_{\rm F}^{*}$, with $T_{\rm F}^{*}$ being   
the Fermi temperature of quasiparticles,  
the $T$ dependence of Im$\Sigma_{\rm f}^{\rm R}({\rm i}\delta)$ is given as 
${\rm Im}\Sigma^{\rm R}_{\tilde{\rm f}}({\rm i}\delta)
\propto -\left({T}/{\eta}\right)^{2}\propto -T^{2(1-\zeta)}$.
where we have used the definition of scaling exponent of valence susceptibility $\eta\propto T^{\zeta}$. 
Then, the $T$ dependence in the resistivity [Eq.\ (\ref{eq:7})] is given as  
\begin{equation}
\rho_{\rm imp}(T)\propto \left[{\rm const.}-T^{2(1-\zeta)}\right].
\label{eq:23}
\end{equation}
This $T$ dependence is to be compared with $\rho(T)\propto ({\rm const.}-T^{n})$ with $n\simeq 0.75$ 
reported in Ref.\ \citen{Shiino}. According to the theoretical prediction,~\cite{Watanabe1} 
the exponent $\zeta$ is weakly $T$ dependent and takes $0.5\le\zeta\lsim0.7$. 
The exponent $n\simeq 0.75$ at $x=0.05$ is reproduced by taking $\zeta\simeq 0.63$. 
{This value of the critical exponent $\zeta$ is actually consistent with the measured 
criticality in the magnetic susceptibility  $\chi\sim T^{-\zeta}$ (with $\zeta\simeq0.67$) at 
$x=0.02$~\cite{Shiino}, considering some uncertainties with inevitable measurement 
errors and fitting to obtain the critical exponents.

\section{Conclusion}
On the basis of a new formulation treating the effect of a diluted system of Ce based heavy 
fermion metals with QCVF, we have succeeded in 
explaining the anomalous non-Fermi liquid $T$ dependence of the resistivity, 
$\rho(T)\propto ({\rm const.}-T^{n})$ with $n\simeq 0.75$, observed  
in {Ce$_{x}$La$_{1-x}$Cu$_{5.62}$Au$_{0.38}$ ($x=0.02-0.10$) in the low temperature region 
{$T<0.5\ $K}~\cite{Shiino}.

\section*{Acknowledgments}
We are grateful to N. K. Sato and T. Shiino for showing us experimental data 
prior to publication. 
This work was supported by JSPS KAKENHI Grant Numbers 18H04326, 18K03542, and 17K05555.

\end{document}